\documentclass[11pt]{article}
\usepackage[auth-sc,affil-it]{authblk}
\usepackage[utf8]{inputenc}
\usepackage{amsmath}
\usepackage{amsthm}
\usepackage{amssymb}
\usepackage{amsfonts}
\usepackage{wasysym}
\usepackage{algorithm}
\usepackage{algpseudocode}
\usepackage{bbm}

\usepackage{tikz}
\usetikzlibrary{calc}
\usetikzlibrary{arrows}
\usetikzlibrary{shadows}
\usetikzlibrary{positioning}
\usetikzlibrary{decorations.pathmorphing,shapes}
\usetikzlibrary{shapes,backgrounds}
\usepackage{mathrsfs}

\newcommand{\ignore}[1]{}
\newcommand{\etal}{{\em et al.}~}

\newtheorem{theorem}{Theorem}

\newtheorem{corollary}[theorem]{Corollary}
\newtheorem{proposition}[theorem]{Proposition}
\newtheorem{example}[theorem]{Example}
\newtheorem{claim}[theorem]{Claim}

\newcommand{\ket}[1]{| #1 \rangle}
\newcommand{\bra}[1]{\langle #1 |}

\newcommand{\ZZ}{\mathbb{Z}}
\newcommand{\RR}{\mathbb{R}}
\newcommand{\CC}{\mathbb{C}}
\newcommand{\HH}{\mathcal{H}}
\newcommand{\mysl}[2]{\mathfrak{sl}_{#1}{#2}}

\newcommand{\one}{\mathbbm{1}}
\newcommand{\norm}[1]{\lVert #1 \rVert}
\newcommand{\ip}[2]{\langle #1, #2 \rangle}

\DeclareMathOperator{\rk}{rank}

\DeclareMathOperator{\spn}{span}
\DeclareMathOperator{\diag}{diag}
\DeclareMathOperator{\myend}{End}

\DeclareMathOperator{\Mat}{Mat}

\newcommand\restr[2]{{
  \left.\kern-\nulldelimiterspace 
  #1 
  \littletaller 
  \right|_{#2} 
  }}
\newcommand{\littletaller}{\mathchoice{\vphantom{\big|}}{}{}{}}

\title{Quantum state transfer in graphs with tails}

\author[1]{Pierre-Antoine Bernard}

\author[2]{Christino Tamon}

\author[1]{Luc Vinet}

\author[3]{Weichen Xie}

\affil[1]{Centre de Recherches Math\'{e}matiques, Universit\'{e} de Montr\'{e}al}
\affil[2]{Department of Computer Science, Clarkson University}
\affil[3]{Department of Mathematics, Clarkson University}

\date{\today}

\begin{document}
\maketitle

\vspace{-.25in}
\begin{abstract}
We consider quantum state transfer on finite graphs which are attached to infinite paths.
The finite graph represents an operational quantum system for performing useful quantum information tasks. 
In contrast, the infinite paths represent external infinite-dimensional systems which have limited (but nontrivial) 
interaction with the finite quantum system. 
We show that {\em perfect} state transfer can surprisingly still occur on the finite graph 
even in the presence of the infinite tails. 
Our techniques are based on a decoupling theorem for eventually-free Jacobi matrices, 
equitable partitions, and standard Lie theoretic arguments. 
Through these methods, we rehabilitate the notion of a dark subspace which had been so far
viewed in an unflattering light.
\end{abstract}

\section{Introduction}

Continuous-time quantum walk on graphs has been fundamental to quantum information and computation.
Bose \cite{b03} studied quantum communication protocols on quantum spin chains using continuous-time
quantum walk (see the comprehensive survey by Kay \cite{k10}). 
On the algorithmic side, Farhi \etal \cite{fgg08} designed a quantum algorithm (with provable speedup)
for a famous Boolean formula evaluation problem. A notable feature of their algorithm is that it employs 
continuous-time quantum walk on finite graphs which are attached to infinite paths (or tails). 
This idea was used subsequently by Childs \cite{c09} to show that continuous-time quantum walk 
is a universal model for quantum computation.

In this work, we explore quantum state transfer (which was the original problem studied by Bose \cite{b03})
on finite graphs which are attached to infinite paths (as considered by Farhi \etal \cite{fgg08}). 
We view the finite graph, on one hand, to represent our internal quantum system for performing basic
quantum information tasks. On the other hand, we view the attached infinite paths as a model of an external 
infinite-dimensional quantum system which has limited (but potentially destructive) interaction with our system. 
The main question which we investigate here is if our quantum protocols can still work in the presence of this 
infinite-dimensional system. We provide an affirmative answer to this question. 

Our main result shows that, under modest conditions, if perfect state transfer occurs in a finite graph, 
then perfect state transfer can still occur in the finite graph even when infinite paths are attached.
The main technique we employ is based on a decoupling theorem for eventually-free Jacobi matrices
as stated by Golinskii \cite{g16}. This technique shows that the notion of a dark subspace is useful 
as a tool to construct {\em protected} computational subspace. This is in contrast to previous works 
which tried to mitigate the existence of these dark subspaces \cite{ldsc18}. 

A second technique we use involves basic representation theory of the Lie algebra $\mysl{2}{\CC}$. 
This allows us to construct {\em perfect} state transfer in the Cartesian square of Krawtchouk chains 
(or quotients of the $n$-cube) and the $n$-cube itself when both are attached to an infinite tail. 
These constructions exhibit perfect state transfer between multiple pairs of quantum states on the 
same graph. Through these, we establish connections between the dark subspaces and the walk modules
of the Terwilliger algebra of the $n$-cube (see \cite{terwilliger}).

Finally, we utilize the dual-rail encoding useful for heralded perfect state transfer in quantum 
spin chains (see \cite{bb05,k10}). This method exploits the presence of an anti-symmetric subspace 
in our graphs with tails. 
We view this method in the context of a Cartesian product a perfect state transfer graph $G$ with $K_2$.  
This allows us to protect any graph with perfect state transfer from an infinite tail probe
by simply ``doubling'' the finite graph.

\section{Preliminaries}

We review some basic notation and terminology which will be used throughout.
The set of integers is denoted $\ZZ$ while $\ZZ^+$ denotes the set of positive integers.

The all-one $n \times m$ matrix is denoted $J_{n,m}$, for $n,m \ge 1$; we simply use $J_n$ if $n = m$.
The identity matrix of order $n$ is denoted $I_n$.
Whenever the dimensions are clear from context, we omit them for brevity.
We use $\one_S$ to denote the characteristic vector of a subset $S$.
The unit vectors are denoted $e_j$ or $e(j)$ (with an implied dimension).
The unit matrix $E_{j,k}$ (which is $e_je_k^T$) is zero everywhere except at entry $(j,k)$ 
where it is $1$ (again with implied dimensions).

We adopt standard asymptotic notation where 
$o_n(f_n)$ denote functions $g_n$ for which $g_n/f_n \rightarrow 0$,
$O_n(f_n)$ denote functions $g_n$ for which $g_n/f_n$ is bounded from above by a constant,
as $n \rightarrow \infty$; see Janson \etal \cite{jlr}.

\subsection{Graphs}

The graphs we study are mainly simple and undirected (unless stated otherwise).
For a graph $G$, we denote its vertex and edge sets as $V(G)$ and $E(G)$, respectively.
The adjacency matrix of $G$ is a matrix $A(G)$ whose $(j,k)$ entry is $1$ if $(j,k) \in E(G)$ and $0$ otherwise.
The neighborhood of a vertex $u$ in $G$ is denoted $N_G(u) = \{v \in V(G) : (u,v) \in E(G)\}$.
The degree $\deg(u)$ of vertex $u$ is the cardinality of $N_G(u)$.
We use standard notation to denote common families of graphs:
$K_n$ for complete graphs (cliques), $\overline{K}_n$ for the empty graphs (cocliques),
$P_n$ for paths, $Q_n$ for the binary $n$-cubes. 
See Godsil and Royle \cite{gr} for further background on graph theory.

Given two graphs $G$ and $H$, the {\em join} $G + H$ is the graph obtained by taking a disjoint union 
of $G$ and $H$ and then adding all edges $(g,h)$ for every $g \in G$ and $h \in H$.
The adjacency matrix of $G+H$ is given by
\[
	A(G+H) =
	\begin{pmatrix}
	A(G) & J_{n,m} \\
	J_{m,n} & A(H)
	\end{pmatrix}
\]
where $G$ has $n$ vertices and $H$ has $m$ vertices. 
The {\em cone} of a graph $G$ is given by $\widehat{G} := K_1 + G$.
By extension, the $m$-cone of $G$ is given by $\overline{K}_m + G$.

The {\em one-sum} of two graphs $G$ and $H$ at a vertex $u$ is obtained by identifying 
a vertex of $G$ with a vertex of $H$; thus, $V(G) \cap V(H) = \{u\}$. The adjacency matrix 
of this one-sum is given by
\[
	\begin{pmatrix}
	A(G \setminus u)& \one_{N_G(u)} & 0 \\
	\one_{N_G(u)}^T & 0 			& \one_{N_H(u)}^T \\
	0 				& \one_{N_H(u)} & A(H \setminus u)
	\end{pmatrix}.
\]

Let $G$ be a graph on $n$ vertices and let $\mathcal{Y} = \{H_i(r_i) : i=1,\ldots,n\}$ be a collection of
rooted graphs (where $r_i$ is a distinguished root vertex in $H_i$).
The {\em rooted product} of $G$ with $\mathcal{Y}$, denoted $G^{\mathcal{Y}}$, is a graph obtained
from the one-sum of $G$ with $H_i$ by identifying vertex $i$ of $G$ with vertex $r_i$ in $H_i$,
for each $i=1,\ldots,n$; see Godsil and McKay \cite{gm78}.

Given a graph $G=(V,E)$, a vertex partition $\pi = (V_j)$ of $G$ is called {\em equitable} if
$V = \bigcup_j V_j$ is a disjoint union, each induced subgraph $G[V_j]$ is $d_j$-regular for some $d_j$, 
and each induced bipartite subgraph $G[V_j,V_k]$ is $(d_{jk},d_{kj})$-biregular for some $d_{jk}$ and $d_{kj}$.
We call each $V_j$ a {\em cell} of the equitable partition $\pi$.
The partition matrix of $\pi$ is given by $Q = (q_{uj})$ where $q_{uj} = \one_{\{u \in V_j\}} \cdot |V_j|^{-1/2}$.
Then, the quotient graph $G/\pi$ is a weighted graph whose adjacency matrix is $A(G/\pi) = Q^\dagger A(G) Q$.
For a fixed vertex $u$ of $G$, the equitable partition relative to $u$ is an equitable partition $\pi$ of $G$
where $\{u\}$ is a cell of $\pi$. In the case where $A(G/\pi)$ is a symmetric tridiagonal matrix (or Jacobi matrix)
we say $\pi$ forms a {\em distance} partition relative to $u$.

\subsection{Linear operators}

We will allow our graphs to be infinite with a countable vertex set (see \cite{m82,mw89,gm88}). 
A main example is the infinite path $P_\infty$ with $\ZZ^+$ as vertices and with edges connecting 
consecutive positive integers.
The adjacency matrix of $P_\infty$ is the {\em free Jacobi} matrix $J_0$ given by
\[
	J_0 =
	\begin{pmatrix}
	0 & 1 &   &        & \\
	1 & 0 & 1 &        & \\
	  & 1 & 0 & 1      & \\
	  &   &   & \vdots & 
	\end{pmatrix}.
\]
A Jacobi matrix $J$ is called {\em eventually-free} if $J_0$ can be obtained from $J$
by the removal of the first $k$ rows and $k$ columns, for some finite $k$.
In this case, we also say $J$ is an extension of $J_0$ (see Golinskii \cite{g16}).

Let $\{G_n\}$ be a family of infinite graphs where each $G_n$ is parameterized by a finite subgraph of size $n$.
As an example, consider the infinite {\em lollipop} graph $L_n$ which is the one-sum of the complete graph $K_n$ 
with the infinite path $P_\infty$ (see Figure \ref{fig:clique-tail}).
We say $G$ is {\em locally finite} if $\deg(u)$ is finite for each $u \in V$.
Let $\deg(G) = \sup\{\deg(u) : u \in V\}$. 

We consider the complex separable Hilbert space $\ell^2(V)$ which is equipped with the inner product 
$\ip{x}{y} = \sum_{u \in V} \overline{x}_u y_u$ for $x,y \in \CC^V$ and an induced norm 
$\norm{x}_2 = \sqrt{\ip{x}{x}}$.
Recall $\ell^2(V)$ consists of vectors $x$ with $\norm{x}_2 < \infty$. 
A complete orthonormal system for $\ell^2(V)$ is given by the standard basis $\{e_v : v \in V\}$.

The adjacency matrix $A$ of $G$ is defined on the basis vectors as 
$Ae_{v} = \sum_{u \in V} a_{u,v}e_{u}$ 
(the sum converges if $G$ is locally finite); 
or, $\ip{e_u}{Ae_v} = a_{u,v}$.
If $\deg(G) < \infty$, then the adjacency matrix $A$ is a bounded self-adjoint operator 
(see Mohar and Woess \cite{mw89}). 
We also say that the adjacency operator of $G$ admits a matrix representation 
relative to the standard basis (see \cite{ag}, section 26).

The {\em spectrum} $\sigma(A)$ of $A$ is the set of all $\lambda \in \CC$ for which
$\lambda I - A$ is not invertible.
The {\em point} spectrum $\sigma_p(A)$ is the set of all $\lambda \in \CC$ for which $\lambda I - A$ is not one-to-one.
Any element of $\sigma_p(A)$ is called an {\em eigenvalue} of $A$.
The {\em continuous} spectrum $\sigma_c(A)$ is the set of all $\lambda \in \CC$ for which $\lambda I - A$ is a one-to-one mapping
of $H$ onto a dense proper subspace of $H$.
If $A$ is bounded and self-adjoint, then $\sigma(A) \subset \RR$ and $\sigma(A) = \sigma_p(A) \cup \sigma_c(A)$;
that is, $A$ has no residual spectrum.
Our primary sources for functional analysis are \cite{ag,rudin}.

\begin{theorem} (Spectral theorem) 
If $A$ is a bounded self-adjoint operator on a complex Hilbert space $H$,
then there exists a unique resolution of the identity $E$ on the Borel subsets of $\sigma(A)$ which satisfies
\[
	A = \int_{\sigma(A)} \lambda \ dE(\lambda).
\]
Moreover, if $f$ is a bounded Borel function on $\sigma(A)$, then there is a linear operator $f(A)$ where
\begin{equation} \label{eqn:spectral-ip}
	\ip{f(A)x}{y} = \int_{\sigma(A)} f(\lambda) dE_{x,y}(\lambda)
\end{equation}
for every $x,y \in H$, where $E_{x,y}(\omega) = \ip{E(\omega)x}{y}$ is a complex measure
over the Borel subsets of $\sigma(A)$. 
\end{theorem}

Based on \eqref{eqn:spectral-ip}, it is customary to adopt the notation
\[
	f(A) = \int_{\sigma(A)} f \ dE.
\]
It is often convenient to assume the resolution $E$ of the identity is defined over 
all Borel subsets of $\CC$ by letting $E(\omega) = 0$ if $\omega \cap \sigma(A) = \emptyset$.

The following result is the basis of our main technique.

\begin{theorem} (see \cite{ag}, Theorem 3, section 40) 
Let $\HH$ be a complex separable Hilbert space and let $A$ be a bounded self-adjoint operator on $\HH$.
Let $W_k$ ($k=1,2,\ldots,m$) be pairwise orthogonal invariant subspaces of $A$; that is,
$\HH = \bigoplus_{k=1}^{m} W_k$ and $AW_k \subset W_k$, for every $k$.
Let $P_k$ be the projection operator on $W_k$ and $A_k$ be the restriction of $A$ to $W_k$.
Then, for every $\psi \in \HH$, we have
\[
	A\psi = \sum_{k=1}^{m} A_kP_k\psi.
\]
\end{theorem}

\subsection{Quantum walk}

Let $G=(V,E)$ be an infinite graph whose adjacency operator $A$ is a bounded self-adjoint linear operator on $\ell^2(V)$. 
A continuous-time quantum walk on $G$ is given by the unitary operator
\[
	e^{-itA} = \int_{\sigma(A)} e^{-it\lambda} \ dE_\lambda.
\]
We say that a graph $G$ has {\em perfect state transfer} (adopting Bose \cite{b03})
between vertices $u$ and $v$ if there is a time $\tau$ so that
\[
	|\ip{e_v}{e^{-i\tau A}e_u}| = 1.
\]
A family $\{G_n\}$ of graphs has {\em asymptotically efficient} perfect state transfer (see Chen \etal \cite{cmf16})
between vertices $u$ and $v$ if there is a time $\tau = n^{O(1)}$ so that
\[
	|\ip{e_v}{e^{-i\tau A(G_n)}e_u}| = 1-o_n(1).
\]

\section{Conical illusion}

\begin{figure}[t]
\begin{center}
\begin{tikzpicture}[
    main node/.style={circle,draw,font=\bfseries}, main edge/.style={-,>=stealth'},
    scale=0.5,
    stone/.style={},
    black-stone/.style={black!80},
    black-highlight/.style={outer color=black!80, inner color=black!30},
    black-number/.style={white},
    white-stone/.style={white!70!black},
    white-highlight/.style={outer color=white!70!black, inner color=white},
    white-number/.style={black}]
\tikzset{every loop/.style={thick, min distance=17mm, in=45, out=135}}


\tikzstyle{every node}=[draw, thick, shape=circle, circular drop shadow, fill={gray}];
\path (+1.25,+2) node (p1) [scale=0.8] {};
\path (-1.25,+2) node (p2) [scale=0.8] {};
\path (-3.0,0) node (p3) [scale=0.8] {};
\path (-1.25,-2) node (p4) [scale=0.8] {};
\path (+1.25,-2) node (p5) [scale=0.8] {};

\tikzstyle{every node}=[draw, thick, shape=circle, circular drop shadow, fill={white}];
\path (+3.0,0) node (p0) [scale=0.8] {};

\path (+5.0,0) node (pa) [scale=0.8] {};
\path (+7.0,0) node (pb) [scale=0.8] {};
\path (+9.0,0) node (pc) [scale=0.8] {};
\path (+11.0,0) node (pd) [scale=0.8] {};
\path (+13.0,0) node (pe) [scale=0.8] {};

\draw[thick]
    (p0) -- (p1) -- (p2) -- (p3) -- (p4) -- (p5) -- (p0);
\draw[thick]
    (p0) -- (p2) -- (p4) -- (p0);
\draw[thick]
    (p1) -- (p3) -- (p5) -- (p1);
\draw[thick]
    (p0) -- (p3);
\draw[thick]
    (p1) -- (p4);
\draw[thick]
    (p2) -- (p5);
\draw[thick]
    (p0) -- (p4) -- (p2) -- (p0);
\draw[thick]
    (p0) -- (p5) -- (p4) -- (p3) -- (p2) -- (p1) -- (p0);

\draw[thick]
    (p0) -- (pa) -- (pb) -- (pc) -- (pd) -- (pe);

\tikzstyle{every node}=[];
\node at (+15,0) {$\ldots$};
\end{tikzpicture}
\vspace{.1in}
\caption{The infinite lollipop graph $L_n$.
All shaded vertices are sedentary.}
\label{fig:clique-tail}
\end{center}
\end{figure}
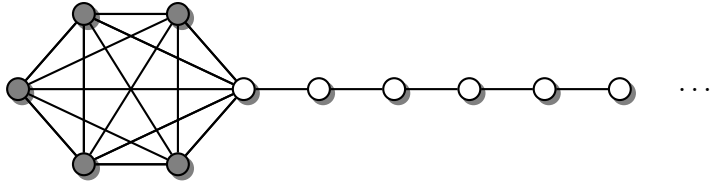

We begin by examining a small yet illustrative example of an infinite graph.
A vertex $u$ in a graph $G_n$ is called {\em sedentary} (see Godsil \cite{g17}) if 
for all time $t$, we have
\[
	|\ip{e_u}{e^{-itA(G_n)}e_u}| = 1-o_n(1).
\]
It is known that the (finite) clique $K_n$ is sedentary at any vertex.
To see this, note that the spectral decomposition of $A(K_n)$ is given by
$A(K_n) = (n-1) J/n - (I - J/n)$, which shows that
\[
	e^{-itA(K_n)} = e^{-it(n-1)} J/n + e^{it}(I - J/n).
\]
This immediately yields
\[
	|\ip{e_u}{e^{-itA(K_n)}e_u}| = |e^{-it(n-1)}/n + e^{it}(1 - 1/n)| = 1 - o_n(1).
\]
We show that the clique is still sedentary even in the presence of an infinite tail 
(see Figure \ref{fig:clique-tail}).

\begin{proposition} \label{prop:clique-tail}
For $n \ge 2$, let $L_n$ be the infinite graph that is the one-sum of the clique $K_n$ 
and the infinite path $P_\infty$ at vertex $u$.
Then, any vertex $v \neq u$ of $K_n$ in $L_n$ is sedentary.

\begin{proof}
Following Golinskii \cite{g16}, we label the vertices of $K_n$ with $1,2,\ldots,n$, where $n$ is the attachment vertex,
and the vertices of $P_\infty$ with the integers $n+1,n+2,\ldots$.
The adjacency matrix of $L_n$ is a bounded self-adjoint operator in $\ell_2(\ZZ^+)$.
Relative to the standard basis $\{e_{k} : k \in \ZZ^+\}$, we have
\[
	A(L_n) = 
	\begin{pmatrix}
	A(K_n) & E_{n,1} \\
	E_{1,n} & J_0
	\end{pmatrix}
\]
where $J_0$ is the free Jacobi matrix.
Now, define a new basis $\{\tilde{e}_k : k \in \ZZ^+\}$ where 
$\tilde{e}_k = e_{k}$, for $k=n,n+1,\ldots$, and
\[
	\tilde{e}_{n-1} = \frac{1}{\sqrt{n-1}}\sum_{k=1}^{n-1} e_{k}
\]
and
\[
	\tilde{e}_{k} = \frac{1}{\sqrt{n-1}}\sum_{j=1}^{n-1} \zeta_{n-1}^{(j-1)k} e_{j}
	\ \hspace{.2in} \
	(k=1,\ldots,n-2)
\]
where $\zeta_p = e^{2\pi i/p}$ denotes the principal $p$-th root of unity.
The action of $A(L_n)$ on the new basis is given by
\[
	U^{-1}A(L_n)U = 
	\begin{pmatrix}
	-\mathbb{I}_{n-2} & O \\
	O & \tilde{J}_0
	\end{pmatrix}
\]
where $\tilde{J}_0$ is an eventually-free Jacobi matrix (of rank two) given by
\begin{equation} \label{eqn:clique-quotient}
	\tilde{J}_0 =
	\begin{pmatrix}
	n-2 		& \sqrt{n-1} 	& O 	\\ 
	\sqrt{n-1} 	& 0 			& e_1^T	\\
	O			& e_1			& J_0 	
	\end{pmatrix}.
\end{equation}
Here, $U$ is the unitary operator which maps $e_k$ to $\tilde{e}_k$ for every $k=1,2,\ldots$.
Thus, 
$W_0 = \spn\{\tilde{e}_{1},\ldots,\tilde{e}_{n-2}\}$ 
and 
$W_\infty = \spn\{\tilde{e}_{n-1},\tilde{e}_{n},\ldots\}$
are orthogonal invariant subspaces which span the entire space.

Finally, to see why any vertex $j \in \{1,\ldots,n-1\}$ in $K_n$ is sedentary, observe that
\[
	\ip{e_j}{e^{-itA(L_n)}e_j} = \frac{e^{it}}{n-1} \sum_{k=1}^{n-2} |\ip{e_j}{\tilde{e}_k}|^2 
		+ O\left(\frac{1}{n-1}\right),
\]
which shows $|\ip{e_j}{e^{-itA(L_n)}e_j}| = 1-o_n(1)$ for all $t$.
\end{proof}
\end{proposition}

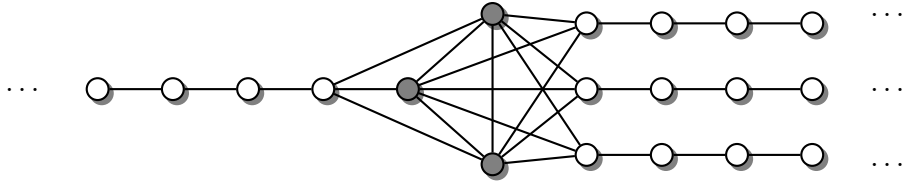
\begin{figure}[t]
\begin{center}
\begin{tikzpicture}[
    main node/.style={circle,draw,font=\bfseries}, main edge/.style={-,>=stealth'},
    scale=0.5,
    stone/.style={},
    black-stone/.style={black!80},
    black-highlight/.style={outer color=black!80, inner color=black!30},
    black-number/.style={white},
    white-stone/.style={white!70!black},
    white-highlight/.style={outer color=white!70!black, inner color=white},
    white-number/.style={black}]
\tikzset{every loop/.style={thick, min distance=17mm, in=45, out=135}}


\tikzstyle{every node}=[draw, thick, shape=circle, circular drop shadow, fill={gray}];
\path (-0.5,+2) node (p2) [scale=0.8] {};
\path (-2.75,0) node (p3) [scale=0.8] {};
\path (-0.5,-2) node (p4) [scale=0.8] {};

\tikzstyle{every node}=[draw, thick, shape=circle, circular drop shadow, fill={white}];
\path (-5,0) node (s0) [scale=0.8] {};
\path (+2,1.75) node (p1) [scale=0.8] {};
\path (+2,0) node (p0) [scale=0.8] {};
\path (+2,-1.75) node (p5) [scale=0.8] {};

\path (-7.0,0) node (sa) [scale=0.8] {};
\path (-9.0,0) node (sb) [scale=0.8] {};
\path (-11.0,0) node (sc) [scale=0.8] {};

\path (+4.0,0) node (pa) [scale=0.8] {};
\path (+6.0,0) node (pb) [scale=0.8] {};
\path (+8.0,0) node (pc) [scale=0.8] {};

\path (+4.0,1.75) node (qa) [scale=0.8] {};
\path (+6.0,1.75) node (qb) [scale=0.8] {};
\path (+8.0,1.75) node (qc) [scale=0.8] {};

\path (+4.0,-1.75) node (ra) [scale=0.8] {};
\path (+6.0,-1.75) node (rb) [scale=0.8] {};
\path (+8.0,-1.75) node (rc) [scale=0.8] {};

\draw[thick]
    (p2) -- (p3);
\draw[thick]
    (p3) -- (p4);
\draw[thick]
    (p4) -- (p2);

\draw[thick]
    (s0) -- (p2);
\draw[thick]
    (s0) -- (p3);
\draw[thick]
    (s0) -- (p4);
\draw[thick]
    (p0) -- (p2);
\draw[thick]
    (p0) -- (p3);
\draw[thick]
    (p0) -- (p4);
\draw[thick]
    (p1) -- (p2);
\draw[thick]
    (p1) -- (p3);
\draw[thick]
    (p1) -- (p4);
\draw[thick]
    (p5) -- (p2);
\draw[thick]
    (p5) -- (p3);
\draw[thick]
    (p5) -- (p4);

\draw[thick]
    (s0) -- (sa) -- (sb) -- (sc);
\draw[thick]
    (p0) -- (pa) -- (pb) -- (pc);
\draw[thick]
    (p1) -- (qa) -- (qb) -- (qc);
\draw[thick]
    (p5) -- (ra) -- (rb) -- (rc);

\tikzstyle{every node}=[];
\node at (-13,0) {$\ldots$};
\node at (+10,2) {$\ldots$};
\node at (+10,0) {$\ldots$};
\node at (+10,-2) {$\ldots$};
\end{tikzpicture}
\vspace{.1in}
\caption{The join $\overline{K}_m + K_n$ attached to multiple tails.
All shaded vertices are still sedentary.
}
\label{fig:clique-many-tail}
\end{center}
\end{figure}

The next result shows that even with an {\em arbitrary} but finite number of infinite tails
the clique is still sedentary. But, we make a slight shift in our approach (which 
sheds light on the title of this section); see Figure \ref{fig:clique-many-tail}.

\begin{theorem} \label{thm:multicone}
Let $n$ and $m$ be positive integers where $n \ge 2$ and $m \ge 1$.
Let $H$ be the infinite graph obtained from the join $\overline{K}_m + K_n$ 
by attaching separate infinite paths to each vertex of the coclique $\overline{K}_m$.
Then, any vertex of the clique $K_n$ is sedentary.

\begin{proof}
The quotient of $H$ is the clique $K_n$ attached to a single infinite path where each vertex
of $K_n$ is connected to the first vertex of $P_\infty$ by an edge of weight $\sqrt{m}$.
The proof now proceeds as in Proposition \ref{prop:clique-tail} without major changes.
\end{proof}
\end{theorem}

Observe that Proposition \ref{prop:clique-tail} is a corollary of Theorem \ref{thm:multicone} when $m=1$.

It seems curious to investigate sedentariness as it is the opposite of quantum state transfer.
Yet, as we will see, the above results reveal a dark subspace which holds interesting properties.

\section{Dark subspace transport over cones}

We explore properties of a finite invariant subspace of graphs with tails.
First, observe that the clique $K_n$ is a cone over a smaller clique, that is, $K_n = K_1 + K_{n-1}$. 
Now, we consider the cone $\widehat{G}_n := K_1 + G_n$ of an arbitrary  $(n,d)$-regular graph $G_n$.
The following result shows that the one-sum of $\widehat{G}_n$ with $P_\infty$ at the conical vertex
inherits the state transfer properties of $G_n$.

\begin{theorem} \label{thm:cone-transport}
Let $G_n$ be a connected $d$-regular graph which admits efficient perfect state transfer between vertices $u$ and $v$ at time $\tau$.
Let $H$ be the infinite graph that is the one-sum of the cone $\widehat{G}_n$ and the infinite path $P_\infty$ at the conical vertex.
Then, $H$ has asymptotically efficient perfect state transfer between $u$ and $v$ at time $\tau$.

\begin{proof}
Suppose the adjacency matrix of $G_n$ is diagonalized by the unitary matrix $Z$, that is,
$Z^\dagger A(G_n)Z = \Lambda$, where $\Lambda = \diag(\lambda_1,\ldots,\lambda_n)$ is the diagonal
matrix of the eigenvalues and the columns of $Z$ are eigenvectors of $A(G_n)$. 
Let $A(G_n) = \sum_{k=1}^{n} \lambda_k z_k z_k^\dagger$ with $\lambda_n = d$ and $z_n = \frac{1}{\sqrt{n}}\one_n$.
Similar to Proposition \ref{prop:clique-tail}, the adjacency matrix of $H$ can be decomposed into
\begin{equation} \label{eqn:factor}
	U^{-1} A(H) U =
	\begin{pmatrix}
	\Lambda_0 & O \\
	O   & \tilde{J}_0
	\end{pmatrix},
	\ \hspace{.2in} \
	\mbox{ where } \ \
	U = 
	\begin{pmatrix}
	Z & O \\
	O & I
	\end{pmatrix}.
\end{equation}
Here, $U$ is the unitary change of basis transformation where $\tilde{e}_j = Ue_j$, $j=1,2,\ldots$, 
shows the new basis vectors in terms of the standard basis vectors. By the definition of $U$, notice 
that $\tilde{e}_j = e_j$ whenever $j \ge n+1$.
Also, $\Lambda_0 = \diag(\lambda_1,\ldots,\lambda_{n-1})$ is a diagonal matrix containing
all non-principal eigenvalues (with multiplicities) of $G_n$.
The eventually-free Jacobi matrix $\tilde{J}_0$ is given by
\[
	\tilde{J}_0 =
	\begin{pmatrix}
	d & \sqrt{n} & O \\
	\sqrt{n} & 0 & e_1^T \\
	O & e_1 & J_0
	\end{pmatrix}.
\]
Let $P$ denote the projection operator onto the subspace $\spn\{\tilde{e}_{j} : j \ge n\}$.
From \eqref{eqn:factor}, we have
\begin{eqnarray*}
\ip{e_v}{e^{-i\tau A(H)}e_{u}} 
	& = & \ip{e_v}{U \exp\left(-i\tau \begin{pmatrix} \Lambda_0 & 0 \\ 0 &  \tilde{J}_0 \end{pmatrix}\right) U^{-1}e_u} \\
	& = & \ip{e_v}{U \begin{pmatrix} e^{-i\tau \Lambda_0} & 0 \\ 0 & e^{-i\tau \tilde{J}_0} \end{pmatrix} U^{-1}e_u}.
\end{eqnarray*}
The last expression equals to
\[
	\ip{e_v}{\sum_{j=1}^{n-1} e^{-i\tau \lambda_j}z_j z_j^T e_u} + \ip{e_v}{UP e^{-i\tau \tilde{J}_0} PU^{-1}e_u}. 
\]
Therefore,
\begin{eqnarray*}
\ip{e_v}{e^{-i\tau A(H)}e_{u}} 
	& = & \ip{e_v}{\sum_{j=1}^{n-1} e^{-i\tau \lambda_j} z_j z_j^\dagger e_u} 
		+ \ip{e_v}{z_n}\ip{z_n}{e^{-i\tau \tilde{J}_0} z_n} \ip{z_n}{e_u} \\
	& = & \ip{e_v}{e^{-i\tau A(G_n)}e_u} - \frac{e^{-i\tau d}}{n} + \frac{1}{n} \bra{z_n}e^{-i\tau\tilde{J}_0}\ket{z_n} \\
	& = & \ip{e_v}{e^{-i\tau A(G_n)}e_u} + O(n^{-1}).
\end{eqnarray*}
This shows $|\ip{e_v}{e^{-i\tau A(H)}e_u}| = 1 - o_n(1)$.
\end{proof}
\end{theorem}

\begin{figure}[t]
\begin{center}
\begin{tikzpicture}[
    main node/.style={circle,draw,font=\bfseries}, main edge/.style={-,>=stealth'},
    scale=0.5,
    stone/.style={},
    black-stone/.style={black!80},
    black-highlight/.style={outer color=black!80, inner color=black!30},
    black-number/.style={white},
    white-stone/.style={white!70!black},
    white-highlight/.style={outer color=white!70!black, inner color=white},
    white-number/.style={black}]
\tikzset{every loop/.style={thick, min distance=17mm, in=45, out=135}}


\tikzstyle{every node}=[draw, thick, shape=circle, circular drop shadow, fill={gray}];
\path (0.95,-3.5) node (p000) [scale=0.8] {};
\path (0.95,+3.5) node (p111) [scale=0.8] {};

\tikzstyle{every node}=[draw, thick, shape=circle, circular drop shadow, fill={white}];
\path (-2,-1.05) node (p100) [scale=0.8] {};
\path (0,-1.35) node (p010) [scale=0.8] {};
\path (+2,-1.5) node (p001) [scale=0.8] {};
\path (-2,+1.05) node (p110) [scale=0.8] {};
\path (0,+1.35) node (p101) [scale=0.8] {};
\path (+2,+1.5) node (p011) [scale=0.8] {};

\path (+5.0,0) node (pa) [scale=0.8] {};
\path (+7.0,0) node (pb) [scale=0.8] {};
\path (+9.0,0) node (pc) [scale=0.8] {};
\path (+11.0,0) node (pd) [scale=0.8] {};
\path (+13.0,0) node (pe) [scale=0.8] {};

\draw[thick]
    (p000) -- (p100) -- (p101) -- (p001) -- (p000);
\draw[thick]
    (p010) -- (p110) -- (p111) -- (p011) -- (p010);
\draw[thick]
    (p000) -- (p010);
\draw[thick]
    (p100) -- (p110);
\draw[thick]
    (p101) -- (p111);
\draw[thick]
    (p001) -- (p011);

\draw[thick]
    (pa) -- (p000)
    (pa) -- (p100)
    (pa) -- (p010)
    (pa) -- (p001)
    (pa) -- (p110)
    (pa) -- (p101)
    (pa) -- (p011)
    (pa) -- (p111);

\draw[thick]
    (pa) -- (pb) -- (pc) -- (pd) -- (pe);

\tikzstyle{every node}=[];
\node at (+15,0) {$\ldots$};
\end{tikzpicture}
\vspace{.1in}
\caption{A cone over the cube $Q_n$ attached to an infinite tail.
Asymptotically efficient perfect state transfer occurs between the shaded antipodal vertices
with fidelity $1 - O(1/2^n)$ at time $(\pi/2)n$.
}
\label{fig:cube-tail}
\end{center}
\end{figure}
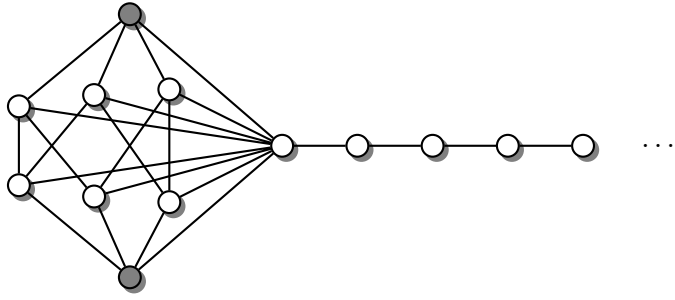

\begin{example}
The one-sum of the cone of the $n$-cube with $P_\infty$ at the conical vertex
has asymptotically efficient perfect state transfer between any pair of antipodal vertices of $Q_n$ 
(see Figure \ref{fig:cube-tail}).
The fidelity of state transfer is $1 - O(1/2^n)$ at the usual time of $(\pi/2)n$.
\end{example}

We remark that Theorem \ref{thm:cone-transport} allows for any other quantum state transfer tasks
(such as, fractional revival) provided the initial and target states have negligible overlap
with the principal eigenvector of the graph.

\section{Decoupling}

Proposition \ref{prop:clique-tail} offers another useful insight.
The eventually-free Jacobi matrix in the infinite lollipop graph 
has the quotient of the clique as its leading principal submatrix.
We generalize this to graphs with well-behaved quotients (for example, distance-regular graphs). 
Our primary goal is still to understand the structure of the underlying protected (dark) subspaces.

Let $G$ be a graph with adjacency matrix $A$ and let $v$ be one of its vertices.
The {\em walk matrix} of $G$ with respect to $v$ is defined as
\[
	W_G(v) = 
	\begin{pmatrix}
	e_v & A e_v & A^2 e_v & \ldots & A^{n-1} e_v
	\end{pmatrix}.
\]
Thus, the columns of the walk matrix generate the Krylov subspace of $A$ relative to the unit vector $e_v$.
The vertex $v$ is called {\em controllable} if $W_G(v)$ has full rank (see Godsil \cite{g12}).

The following result appeared in Golinskii \cite{g16}.
We will provide an alternative proof which illuminates the structure of the dark (protected) subspace.

\begin{theorem} \label{thm:factor}
Let $G_n$ be a connected graph and let $v$ be one of its vertices. 
Let $H$ be the infinite graph that is the one-sum of a finite graph $G_n$ with the infinite path $P_\infty$ at vertex $v$. 
Then, there is an invertible matrix $U$ so that 
\[
	U^{-1}A(H)U = \begin{pmatrix} \Im_v & 0 \\ 0 & \tilde{J}_0 \end{pmatrix} 
\]
where $\dim(\Im_v) = n - \rk(W_{G_n}(v))$ and $\tilde{J}_0$ is an eventually-free Jacobi matrix.

Moreover, 
if $G_n$ admits an equitable distance partition $\pi$ relative to $v$ so that $A(G_n/\pi)$ is a Jacobi matrix,
then the leading principal submatrix of $\tilde{J}_0$ is $A(G_n/\pi)$.

\begin{proof}
For brevity, let $A = A(G_n)$.
First, we show that we can construct an extension $\tilde{J}_0$ of the free Jacobi matrix $J_0$.
We apply the Lanczos algorithm for tridiagonalizing a symmetric matrix starting with $e_v$
(see Horn and Johnson \cite{hj13}, 3.5P5, page 221).
This algorithm will find an orthonormal basis for the Krylov subspace $K = \{e_v,A e_v,\ldots,A^{n-1} e_v\}$.
Assume that $m = \dim K$ and let $\{\tilde{e}_0,\ldots,\tilde{e}_{m-1}\}$ be an orthonormal basis of $K$, 
where $\tilde{e}_0 = e_v$.
Denote $\tilde{e}_{-1}$ as the unit vector corresponding to the unique neighbor of $v$ on the infinite path.
Then, 
\[
	A\tilde{e}_j = \spn\{\tilde{e}_{j-1},\tilde{e}_j,\tilde{e}_{j+1}\}
	\ \hspace{.2in} \ (j=0,\ldots,m-2)
\]
and
\[
	A\tilde{e}_{m-1} = \spn\{\tilde{e}_{m-2},\tilde{e}_{m-1}\}.
\]
This shows that the subspace spanned by $\tilde{e}_0,\ldots,\tilde{e}_{m-1}$ and the standard basis 
$\{e_w : w \in V(P_\infty)\}$ for $P_\infty$ is an invariant subspace of $H$;
moreover, the restriction of $H$ on this subspace forms an extension $\tilde{J}_0$ of $J_0$ 
(because of the tridiagonal structure imposed by the Krylov subspace).

It remains to show that the leading $m \times m$ principal submatrix of $\tilde{J}_0$ is $A(G_n/\pi)$.
By our assumption on $\pi$, we may write $A(G_n)$ in a tridiagonal block form:
\[
	A(G_n) =
	\begin{pmatrix}
	A_1 & B_1^T &		&  		&			&			\\
	B_1 & A_2 	& B_2^T & 		&			&			\\
		& B_2	& A_3	& B_3^T	&			&			\\
		&		&		& \vdots&			&			\\
		&		&		&		& A_{m-1}	& \one_{N_{G_n}(v)}	\\
		&		&		&		& \one_{N_{G_n}(v)}^T	& 0
	\end{pmatrix}.
\]
Here, the last row and column correspond to vertex $v$.
Notice that each $A_i$ is the adjacency matrix of a regular subgraph of $G_n$
and each $B_i$ represents the incidence matrix between subgraphs $A_i$ and $A_{i+1}$,
where $B_i$ has constant row-sum and constant column-sum.
Because of these properties, it is easy to see that the Lanczos algorithm, starting with $e_v$
(last row and column), will construct a basis $\{\tilde{e}_0=e_v,\tilde{e}_1,\ldots,\tilde{e}_{m-1}\}$
so that the action of $H$ on this basis is given by
\[
	\begin{pmatrix}
	d_1 	& d_{1,2} 	& 			&			&	\\
	d_{2,1}	& d_2		& d_{2,3}	&			&	\\
			& d_{3,2}	& d_{3}		& d_{3,4}	&	\\
			&			& \vdots	&			&	\\
			&			&			& d_{m-1}	& 1 \\
			&			&			& |N(v)|	& 0
	\end{pmatrix}
\]
where $d_i$ is the degree of regular subgraph $G[V_i]$ and $d_{i,j}$ denotes the number of neighbors
in $V_{j}$ of each vertex in $V_i$.
After a standard transformation, the previous matrix is equal to the adjacency matrix of 
the symmetrized quotient of $G_n/\pi$.
\end{proof}
\end{theorem}

We refer to the finite invariant subspace of $H$ in Theorem \ref{thm:factor} as the {\em protected (or dark) subspace} of $H$.
Using the above theorem, we observe that a random graph offers no protection.

\begin{corollary} \label{cor:random}
Let $G_n = G(n,1/2)$ be a random graph.
Then, the one-sum of $\widehat{G}_n = K_1 + G_n$ with $P_\infty$ has no protected subspace almost surely.

\begin{proof}
This follows since a random graph $G(n,1/2)$ is controllable relative to $\one_n$ almost surely 
(see O'Rourke and Touri \cite{ot16}).
\end{proof}
\end{corollary}

\section{Dark subspace transport via equitable partitions}

Using Theorem \ref{thm:factor},
we explore the dark subspaces induced by regular subgraphs within the (equitable) distance partitions of graphs.
Our goal is to show that state transfer occurs within each local subgraph under suitable conditions.

Given a regular graph $G$, consider the graph $P_m(G)$ obtained by connecting $m$ disjoint copies of $G$ 
in a series using the graph join operator. So, the adjacency matrix of this graph is $I_m \otimes A(G) + A(P_m) \otimes J$.
We now generalize this {\em series graph} by allowing $m$ (possibly distinct) regular graphs, $G_0,G_1,\ldots,G_{m-1}$ 
and denote the resulting graph as $P_m(G_0,\ldots,G_{m-1})$.

\newcommand{\GG}{\mathcal{G}}

\begin{theorem} \label{thm:path-join}
Let $\GG_n = P_m(\{u\},G_1,\ldots,G_{m-1})$ be a series graph.
Let $H$ be the infinite graph that is the one-sum of $\GG_n$ and the infinite path $P_\infty$ at $u$. 
If there is $G_j$, $j=1,\ldots,m-1$, with perfect state transfer between its vertices $a$ and $b$ at time $\tau$,
then $H$ has asymptotic perfect state transfer between $a$ and $b$ at time $\tau$.

\begin{proof}
We prove the claim by induction on $m$.
The base case of $m=2$ is simply Theorem \ref{thm:cone-transport}.
Next, we prove the case for $m=3$ (and leave the full inductive argument as a straightforward exercise).
Let $\GG_n = P_3(\{u\},G_1,G_2)$ where $G_i$ is a $(n_i,d_i)$-regular graph.
The adjacency matrix $A$ of $\GG_n$ is given by
\[
	A =
	\begin{pmatrix}
	A_2	& J 			& 0 \\
	J 	& A_1 		 	& \one_{V_1} \\
	0 	& \one_{V_1}^T  & 0
	\end{pmatrix}.
\]
By Theorem \ref{thm:factor}, we know that
\[
	A =
	\begin{pmatrix}
	\Im_u & O \\
	O & \tilde{J}_0
	\end{pmatrix},
	\ \hspace{.2in} \mbox{ where } \
	\tilde{J}_0 = 
	\begin{pmatrix}
	d_2 & \sqrt{n_1 n_2} & 0 \\
	\sqrt{n_1 n_2} & d_1 & \sqrt{n_1} \\
	0 & \sqrt{n_1} & 0
	\end{pmatrix}
\]
Suppose $z_1$ is an eigenvector of $A_1$ corresponding to a non-principal eigenvalue $\lambda_1$ 
(that is, $\lambda_1 < d_1$) and 
$z_2$ is an eigenvector of $A_2$ corresponding to a non-principal eigenvalue $\lambda_2$ (that is, $\lambda_2 < d_2$). 
Then,
\[
	(\lambda_1 - A)\begin{pmatrix} 0 \\ z_1 \\ 0 \end{pmatrix} = 0,
	\ \hspace{.2in} \
	(\lambda_2 - A)\begin{pmatrix} z_2 \\ 0 \\ 0 \end{pmatrix} = 0.
\]
This implies that $\Im_u$ is block diagonal.
Therefore, for some invertible matrix $U$, we have
\[
	U^{-1} A U =
	\begin{pmatrix}
	\Im_1 & O & O \\
	O & \Im_2 & O \\
	O & O & \tilde{J}_0
	\end{pmatrix}
\]
So, asymptotic perfect state transfer occurs within each block 
(by a similar argument to Theorem \ref{thm:cone-transport}).
\end{proof}
\end{theorem}

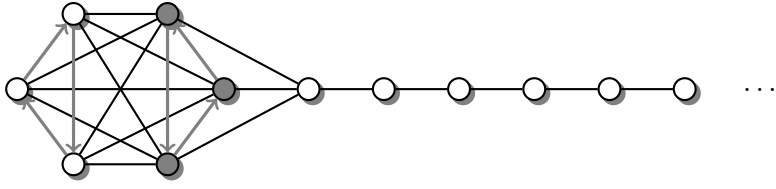
\begin{figure}[t]
\begin{center}
\begin{tikzpicture}[
    main node/.style={circle,draw,font=\bfseries}, main edge/.style={-,>=stealth'},
    scale=0.5,
    stone/.style={},
    black-stone/.style={black!80},
    black-highlight/.style={outer color=black!80, inner color=black!30},
    black-number/.style={white},
    white-stone/.style={white!70!black},
    white-highlight/.style={outer color=white!70!black, inner color=white},
    white-number/.style={black}]
\tikzset{every loop/.style={thick, min distance=17mm, in=45, out=135}}


\tikzstyle{every node}=[draw, thick, shape=circle, circular drop shadow, fill={gray}];
\path (+1.25,+2) node (p100) [scale=0.8] {};
\path (+2.75,0) node (p010) [scale=0.8] {};
\path (+1.25,-2) node (p001) [scale=0.8] {};

\tikzstyle{every node}=[draw, thick, shape=circle, circular drop shadow, fill={white}];
\path (-1.25,+2) node (p110) [scale=0.8] {};
\path (-2.75,0) node (p101) [scale=0.8] {};
\path (-1.25,-2) node (p011) [scale=0.8] {};

\tikzstyle{every node}=[draw, thick, shape=circle, circular drop shadow, fill={white}];
\path (+5,0) node (p000) [scale=0.8] {};

\path (+7.0,0) node (pa) [scale=0.8] {};
\path (+9.0,0) node (pb) [scale=0.8] {};
\path (+11.0,0) node (pc) [scale=0.8] {};
\path (+13.0,0) node (pd) [scale=0.8] {};
\path (+15.0,0) node (pe) [scale=0.8] {};

\draw[thick]
    (p000) -- (p100)
	(p000) -- (p010)
	(p000) -- (p001);

\draw[thick]
	(p100) -- (p110)
	(p100) -- (p101)
	(p100) -- (p011);
\draw[thick]
	(p010) -- (p110)
	(p010) -- (p101)
	(p010) -- (p011);
\draw[thick]
	(p001) -- (p110)
	(p001) -- (p101)
	(p001) -- (p011);

\draw[very thick,->,gray]
	(p100) -- (p001);
\draw[very thick,->,gray]
	(p001) -- (p010);
\draw[very thick,->,gray]
	(p010) -- (p100);

\draw[very thick,->,gray]
	(p110) -- (p011);
\draw[very thick,->,gray]
	(p011) -- (p101);
\draw[very thick,->,gray]
	(p101) -- (p110);

\draw[thick]
    (pa) -- (p000);

\draw[thick]
    (pa) -- (pb) -- (pc) -- (pd) -- (pe);

\tikzstyle{every node}=[];
\node at (+17,0) {$\ldots$};
\end{tikzpicture}
\vspace{.1in}
\caption{A complex Hermitian clique $K_6$ attached to the infinite path $P_\infty$.
The shaded vertices have asymptotically efficient universal perfect state tranfer (PST between every pair).
}
\label{fig:factor-oriented}
\end{center}
\end{figure}

\begin{example}
Let $\vec{K}_3$ be the complex oriented clique of order $3$ whose adjacency matrix is 
\begin{equation} 
	\begin{pmatrix}
	0 & -i & i \\
	i & 0 & -i \\
	-i & i & 0
	\end{pmatrix}.
\end{equation}
We consider the graph $H = P_3(K_1,\vec{K}_3,\vec{K}_3)$ (see Figure \ref{fig:factor-oriented}).
By Theorem \ref{thm:path-join} and an observation in Cameron \etal \cite{cfghst14}, we may conclude
that each local $\vec{K}_3$ has asymptotically efficient {\em universal} perfect state transfer.
\end{example}

\section{Perfect state transfer in a dark subspace}

In this section, we show that {\em perfect} state transfer (with unit fidelity) can be engineered 
to occur in graphs with tails. 
Our techniques are primarily based on the representation theory of the Lie algebra $\mysl{2}{\CC}$ 
and the dual-rail encoding scheme in spin chains.

We recall that the Cartesian product of two graphs $G_1$ and $G_2$, denoted $G_1 \Box G_2$,
is a graph with vertex set $V(G_1) \times V(G_2)$ where the pairs $(a_1,a_2)$ and $(b_1,b_2)$
are adjacent if either $a_1 = b_1$ and $(a_2,b_2) \in E(G_2)$ or $(a_1,b_1) \in E(G_1)$ and $a_2=b_2$.
The adjacency matrix of $G_1 \Box G_2$ is given by $A(G_1) \otimes I + I \otimes A(G_2)$. 
If the two graphs are equal, we use $G^{\Box 2}$ to denote the Cartesian square.

\subsection{Walk modules}

The following example shows perfect {\em pair} state transfer (see Chen and Godsil \cite{cg20})
on $P_3^{\Box 2}$ in the presence of an infinite tail.

\newcommand{\eket}[1]{e_{#1}}
\newcommand{\ebra}[1]{e_{#1}^T}

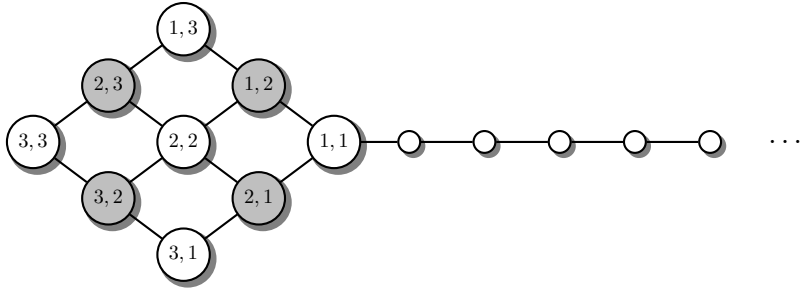
\begin{figure}[t]
\begin{center}
\begin{tikzpicture}[
    main node/.style={circle,draw,font=\bfseries}, main edge/.style={-,>=stealth'},
    scale=0.5,
    stone/.style={},
    black-stone/.style={black!80},
    black-highlight/.style={outer color=black!80, inner color=black!30},
    black-number/.style={white},
    white-stone/.style={white!70!black},
    white-highlight/.style={outer color=white!70!black, inner color=white},
    white-number/.style={black}]
\tikzset{every loop/.style={thick, min distance=17mm, in=45, out=135}}


\tikzstyle{every node}=[draw, thick, shape=circle, circular drop shadow, fill={white}];
\path (-3.0,0) node (p33) [scale=0.7] {$3,3$};

\tikzstyle{every node}=[draw, thick, shape=circle, circular drop shadow, fill={gray!50}];
\path (-1.0,+1.5) node (p23) [scale=0.7] {$2,3$};
\path (-1.0,-1.5) node (p32) [scale=0.7] {$3,2$};

\tikzstyle{every node}=[draw, thick, shape=circle, circular drop shadow, fill={white}];
\path (+1.0,+3.0) node (p13) [scale=0.7] {$1,3$};
\path (+1.0,0) node (p22) [scale=0.7] {$2,2$};
\path (+1.0,-3.0) node (p31) [scale=0.7] {$3,1$};

\tikzstyle{every node}=[draw, thick, shape=circle, circular drop shadow, fill={gray!50}];
\path (3.0,+1.5) node (p12) [scale=0.7] {$1,2$};
\path (3.0,-1.5) node (p21) [scale=0.7] {$2,1$};

\tikzstyle{every node}=[draw, thick, shape=circle, circular drop shadow, fill={white}];
\path (+5,0) node (p11) [scale=0.7] {$1,1$};

\path (+7.0,0) node (pa) [scale=0.8] {};
\path (+9.0,0) node (pb) [scale=0.8] {};
\path (+11.0,0) node (pc) [scale=0.8] {};
\path (+13.0,0) node (pd) [scale=0.8] {};
\path (+15.0,0) node (pe) [scale=0.8] {};

\draw[thick]
    (p11) -- (p12) -- (p13)
	(p11) -- (p21) -- (p31)
	(p12) -- (p22) -- (p32)
	(p21) -- (p22) -- (p23)
	(p13) -- (p23) -- (p33)
	(p31) -- (p32) -- (p33);

\draw[thick]
    (pa) -- (p11);

\draw[thick]
    (pa) -- (pb) -- (pc) -- (pd) -- (pe);

\tikzstyle{every node}=[];
\node at (+17,0) {$\ldots$};
\end{tikzpicture}
\vspace{.1in}
\caption{The {\em fly-swatter} graph: $P_3^{\Box 2}$ attached to an infinite tail.
{\em Perfect} state tranfer occurs between $\frac{1}{\sqrt{2}}(\eket{1,2} - \eket{2,1})$ 
and $\frac{1}{\sqrt{2}}(\eket{2,3}-\eket{3,2})$ at time $\pi/\sqrt{2}$.
}
\label{fig:grid-tail}
\end{center}
\end{figure}

\begin{example}
Suppose the vertex set of $P_3$ is $\{1,2,3\}$ where $1$ and $3$ are the leaves.
Let $H$ be the one-sum of $P_3^{\Box 2}$ and $P_\infty$ at vertex $(1,1)$ (see Figure \ref{fig:grid-tail}).
Let $e_{j,k}$ denote $e_j \otimes e_k$.
We show there is perfect state transfer between the states
$\tfrac{1}{\sqrt{2}}(e_{1,2}-e_{2,1})$ and $\tfrac{1}{\sqrt{2}}(e_{2,3} - e_{3,2})$
at time $\pi/\sqrt{2}$. 
Consider a new basis for $A(H)$. Let $\tilde{e}_k = \eket{k}$, $k \ge 10$, correspond 
to vertices of $P_\infty$, and let
\begin{align*}
\tilde{e}_9 &= \eket{1,1} 
	& \tilde{e}_1 &= \tfrac{1}{\sqrt{3}}(\eket{1,3} - \eket{2,2} + \eket{3,1}) \\
\tilde{e}_8 &= \tfrac{1}{\sqrt{2}}(\eket{1,2} + \eket{2,1}) 
	& \tilde{e}_2 &= \tfrac{1}{\sqrt{2}}(\eket{1,2} - \eket{2,1}) \\
\tilde{e}_7 &= \tfrac{1}{\sqrt{6}}(\eket{1,3} + 2\eket{2,2} + \eket{3,1}) 
	& \tilde{e}_3 &= \tfrac{1}{\sqrt{2}}(\eket{1,3} - \eket{3,1}) \\
\tilde{e}_6 &= \tfrac{1}{\sqrt{2}}(\eket{2,3} + \eket{3,2}) 
	& \tilde{e}_4 &= \tfrac{1}{\sqrt{2}}(\eket{2,3} - \eket{3,2}) \\
\tilde{e}_5 &= \eket{3,3} 
\end{align*}
be the basis vectors for the finite graph.
The subspaces $W_0 = \spn\{\tilde{e}_1,\tilde{e}_2,\tilde{e}_3,\tilde{e}_4\}$
and $W_1 = \spn\{\tilde{e}_k : k \ge 5\}$ are orthogonal and invariant under $A(H)$.
Under this new basis, we have
\[
	A(H) = 
	\begin{pmatrix}
	0 &   &   &   & \\
	  &   & 1 &   & \\
	  & 1 &   & 1 & \\
	  &   & 1 &   & \\
	  &   &   &   & \tilde{J}_0 
	\end{pmatrix}
\]
where $\tilde{J}_0$ is an eventually-free Jacobi matrix.
As the action of $A(H)$ on the invariant subspace spanned by $\{\tilde{e}_2,\tilde{e}_3,\tilde{e}_4\}$
is equivalent to the adjacency matrix of $P_3$, it follows that perfect state transfer occurs 
between $\tilde{e}_2$ and $\tilde{e}_4$ at time $\pi/\sqrt{2}$.
\ignore{
\[
	\tilde{J}_0 = 
	\begin{pmatrix}
	         & \sqrt{2} &          &          &          &   &        & \\
	\sqrt{2} &          & \sqrt{3} &          &          &   &        & \\
	         & \sqrt{3} &          & \sqrt{3} &          &   &        & \\
	         &          & \sqrt{3} &          & \sqrt{2} &   &        & \\
	         &          &          & \sqrt{2} &          & 1 &        & \\
	         &          &          &          & 1        &   & 1      & \\
			 &          &          &          &          & 1 &        & \ldots \\
			 &          &          &          &          &   & \vdots & 
	\end{pmatrix}
\]
}
\end{example}

We now generalize the above example.
For the $n$-cube $Q_n$, let $\pi$ denote the equitable partition whose cells correspond to the
set of vertices with the same Hamming weight. The quotient graph $Q_n/\pi$ is a weighted path
of length $n+1$ (also known as the Krawtchouk chain; see Christandl \etal \cite{cddekl05}).

As we will use some basic Lie theoretic arguments, we briefly recall some relevant terminology;
see \cite{hall,fh} for more background on Lie theory.
The Lie algebra $\mysl{2}{\CC}$ is the algebra of complex $2 \times 2$ traceless matrices
that is equipped with a {\em bracket}\footnote{The Lie bracket is bilinear, anti-symmetric,
and satisfies $[A,[B,C]]+ [B,[C,A]] + [C,[A,B]]=0$.} operation $[A,B] = AB-BA$.
A representation of $\mysl{2}{\CC}$ is a homomorphism $\rho: \mysl{2}{\CC} \rightarrow \myend(V)$ 
for some vector space $V$ of finite dimension (here, $\myend(V)$ is equipped with the natural 
bracket).

\newcommand{\kr}[1]{\widetilde{P}_{#1}}

\begin{theorem} \label{thm:clebsch-gordan}
For any integer $n \ge 2$, let $H_n$ be the infinite weighted graph that is the
one-sum of the Cartesian square $(Q_n/\pi)^{\Box 2}$ and the infinite path $P_\infty$ at vertex $(0_n,0_n)$.
Then, there exist multiple perfect state transfer on $H_n$ at time $\pi/2$.

\begin{proof}
The Clebsch-Gordan decomposition (see \cite{hall}, appendix C) states that 
if $\rho_m$ is an irreducible representation of $\mysl{2}{\CC}$ of dimension $m+1$,
then a tensor product of two irreducible representations can be decomposed into a direct sum of
irreducible representations. In particular, for any $n \ge 1$:
\begin{equation} \label{eqn:clebsch-gordan}
	\rho_n(M) \otimes I + I \otimes \rho_n(M) \ \cong \ \bigoplus_{k=0}^{n} \rho_{2n-2k}(M),
	\ \ \
	\forall M \in \mysl{2}{\CC}.
\end{equation}
To use \eqref{eqn:clebsch-gordan}, we observe there exists $M \in \mysl{2}{\CC}$ 
where $\rho(M) \cong A(Q_{m}/\pi)$, for every $m \ge 2$. 
Recall that a basis for $\mysl{2}{\CC}$ is given by
\begin{align*}
	X &= \begin{pmatrix} 0 & 1 \\ 0 & 0 \end{pmatrix},
	&
	Y &= \begin{pmatrix} 0 & 0 \\ 1 & 0 \end{pmatrix},
	&
	H &= \begin{pmatrix} 1 & 0 \\ 0 & -1 \end{pmatrix}.
\end{align*}
If we let $M = X+Y$, then $\rho(M) = \rho(X) + \rho(Y)$. 
Under the basis of all eigenvectors of $\rho(H)$, we have
\begin{align*}
	\rho(Y) = 
	\begin{pmatrix}
	0 &   &        &   & \\
	1 & 0 &        &   & \\
	  & 1 & 0      &   & \\
	  &   & \vdots &   & \\
	  &   &        & 1 & 0 
	\end{pmatrix},
	\ \
	\rho(X) =
	\begin{pmatrix}
	0 & m &        &        & \\
	  & 0 & 2(m-1) &        & \\
	  &   & 0      & 3(m-2) & \\
	  &   &        & \vdots & \\
	  &   &        &        & m \\
	  &   &        &        & 0
	\end{pmatrix}.
\end{align*}
This shows that $\rho(M) \cong A(Q_m/\pi)$ by a standard basis change.
Hence, perfect state transfer occurs in each of the Krawtchouk chains in \eqref{eqn:clebsch-gordan}.
\end{proof}
\end{theorem}

Next, we show that the $n$-cube itself can be ``immunized'' against the infinite tail.
Recall that $\zeta_n = \exp(2\pi i/n)$ is the primitive $n$th root of unity.
It will be convenient to identify the vertices of the $n$-cube $Q_n$ 
with subsets of $\{1,2,\ldots,n\}$.

\begin{figure}[t]
\begin{center}
\begin{tikzpicture}[
    main node/.style={circle,draw,font=\bfseries}, main edge/.style={-,>=stealth'},
    scale=0.5,
    stone/.style={},
    black-stone/.style={black!80},
    black-highlight/.style={outer color=black!80, inner color=black!30},
    black-number/.style={white},
    white-stone/.style={white!70!black},
    white-highlight/.style={outer color=white!70!black, inner color=white},
    white-number/.style={black}]
\tikzset{every loop/.style={thick, min distance=17mm, in=45, out=135}}


\tikzstyle{every node}=[draw, thick, shape=circle, circular drop shadow, fill={white}];
\path (-4.0,0) node (p111) [scale=0.7] {$111$};

\tikzstyle{every node}=[draw, thick, shape=circle, circular drop shadow, fill={gray!50}];
\path (-1.0,2.5) node (p110) [scale=0.7] {$110$};
\path (-1.0,0) node (p101) [scale=0.7] {$101$};
\path (-1.0,-2.5) node (p011) [scale=0.7] {$011$};

\tikzstyle{every node}=[draw, thick, shape=circle, circular drop shadow, fill={gray!50}];
\path (2.0,+2.5) node (p100) [scale=0.7] {$100$};
\path (2.0,0.0) node (p010) [scale=0.7] {$010$};
\path (2.0,-2.5) node (p001) [scale=0.7] {$001$};

\tikzstyle{every node}=[draw, thick, shape=circle, circular drop shadow, fill={white}];
\path (+5,0) node (p000) [scale=0.7] {$000$};

\path (+7.0,0) node (pa) [scale=0.8] {};
\path (+9.0,0) node (pb) [scale=0.8] {};
\path (+11.0,0) node (pc) [scale=0.8] {};
\path (+13.0,0) node (pd) [scale=0.8] {};
\path (+15.0,0) node (pe) [scale=0.8] {};

\draw[thick]
    (p111) -- (p110)
    (p111) -- (p101)
    (p111) -- (p011)
	(p110) -- (p100)
	(p110) -- (p010)
	(p101) -- (p100)
	(p101) -- (p001)
	(p011) -- (p010)
	(p011) -- (p001)
	(p000) -- (p100)
	(p000) -- (p010)
	(p000) -- (p001);

\draw[thick]
    (pa) -- (p000);

\draw[thick]
    (pa) -- (pb) -- (pc) -- (pd) -- (pe);

\tikzstyle{every node}=[];
\node at (+17,0) {$\ldots$};
\end{tikzpicture}
\vspace{.1in}
\caption{The $3$-cube attached to an infinite tail.
{\em Perfect} state tranfer occurs between 
$\frac{1}{\sqrt{3}}(\eket{100} + \zeta \eket{010} + \zeta^2 \eket{001})$ 
and 
$\frac{1}{\sqrt{3}}(\eket{011} + \zeta \eket{101} + \zeta^2 \eket{110})$ 
at time $\pi/2$,
where $\zeta = \exp(2\pi i/3)$ is the primitive cube root of unity.
}
\label{fig:module-tail}
\end{center}
\end{figure}

\begin{theorem} \label{thm:dark-cube}
For integer $n \ge 2$, let $H_n$ be the infinite graph that is the one-sum of the $n$-cube $Q_n$ 
and the infinite path $P_\infty$ at vertex $0_n$. 
Then, perfect state transfer occurs in $H_n$ between the states
\[
	\frac{1}{\sqrt{n}} \sum_{j=1}^{n} \zeta_n^{j-1} e(\{j\})
	\ \ \mbox{ and } \ \
	\frac{1}{\sqrt{n}} \sum_{j=1}^{n} \zeta_n^{j-1} e([n] \setminus \{j\})
\]
at time $\pi/2$ 

\begin{proof}
By Theorem \ref{thm:factor}, the infinite invariant subspace of $A=A(H)$ is
the walk module $U_0$ generated by $e_\emptyset$ (which corresponds to the
vertex $0_n$). Let $W_d$ be the set of vertices of $Q_n$ of Hamming weight $d$.
Thus, 
\begin{align*}
	U_0 &= \spn\{e(\emptyset), Ae(\emptyset), A^2e(\emptyset), \ldots\} \\
		&= \spn(\{e_\emptyset,\one_{W_1},\one_{W_2},\ldots,\one_{W_n}\} \cup \{e_u : u \in V(P_\infty)\}).
\end{align*}
The action of $A$ on $U_0$ is given by the eventually-free Jacobi matrix
\[
	\restr{A}{U_0} = 
	\begin{pmatrix}
	0      & n      &        &        &   &        &   & \\
	1      & 0      & n-1    &        &   &        &   & \\ 
	       & 2      & 0      &        &   &        &   & \\
	       & \vdots & \vdots & \vdots &   &        &   & \\
	       &        &        & 0      & 1 &        &   & \\
	       &        &        & n      & 0 & 1      &   & \\
	       &        &        &        & 1 & 0      & 1 & \ldots \\
	       &        &        &        &   & \vdots & \vdots & 
	\end{pmatrix}.
\]
Next, consider a state supported on the singleton subsets of $Q_n$ which is defined as
\[
	\psi_1 := \sum_{j=1}^{n} \zeta_n^{j-1} e(\{j\}).
\]
and its walk module
\[
	U_1 = \spn\{\psi_1, A\psi_1, A^2\psi_1, \ldots\}.
\]
It is clear that $U_1$ is an invariant subspace of $A$ and that $U_1 \subset U_0^\perp$.
The latter follows as $A$ is self-adjoint and $\psi_1$ is orthogonal to $\one_{W_1}$.
We now show that $\dim(U_1) = n-1$ and
\begin{equation} \label{eqn:krawtchouk-chain}
	\restr{A}{U_1} \ \cong \
	\begin{pmatrix}
	  & n-2 &        &        & \\
	1 &     & n-3    &        & \\
	  & 2   &        & \ldots & \\
	  &     & \vdots &        & \\
	  &     &        &        & 1 \\
	  &     &        & n-2    &  
	\end{pmatrix}.
\end{equation}
To see this, consider the set $\{\psi_1,\psi_2,\ldots,\psi_{n-1}\}$
of orthogonal basis vectors for $U_1$, where
\[
	\psi_k = \sum_{\substack{S \subset [n]\\|S|=k}} \alpha(S) e(S),
	\ \ \mbox{ where } \ \
	\alpha(S) = \sum_{j \in S} \zeta_n^j.
\]
Then, for $k=1,\ldots,n-2$, we have
\begin{align*}
	\ip{\psi_{k+1}}{A\psi_k} &= k \\
	\ip{\psi_k}{A\psi_{k+1}} &= n-k,
\end{align*}
which shows \eqref{eqn:krawtchouk-chain}.
Thus, it follows $\restr{A}{U_1} \cong A(Q_{n-2}/\pi)$, where $\pi$ is the equitable partition
whose cells are the vertices with the same Hamming weight.
Hence, perfect state transfer occurs in $U_1$ from $n^{-1/2}\psi_1$ 
to $n^{-1/2}\psi_{n-1}$ at time $\pi/2$ as antipodal perfect state transfer
occurs in $A(Q_{n-2}/\pi)$ at that time.
\end{proof}
\end{theorem}

In the proof of Theorem \ref{thm:dark-cube}, the infinite invariant subspace $U_0$ 
is related to the {\em primary module} of the Terwilliger algebra of the $n$-cube
(see Terwilliger \cite{terwilliger}, Definition 7.2). Our dark subspaces correspond 
to the irreducible modules orthogonal to this primary module.

We offer an alternative proof of Theorem \ref{thm:dark-cube} which 
emphasizes the underlying connection with the Lie algebra $\mysl{2}{\CC}$.

\begin{theorem} \label{thm:dark-cube2}
For integer $n \ge 2$, let $H_n$ be the infinite graph that is the one-sum of the $n$-cube $Q_n$ 
and the infinite path $P_\infty$ at vertex $0_n$. 
Then, multiple perfect state transfer occur in $H_n$.

\begin{proof}
Given the $n$-cube $Q_n$, recall the standard lowering $L_n$ and raising $R_n$ operators 
on the lattice of subsets of $\{1,2,\ldots,n\}$. For subsets $A,B \subset \{1,2,\ldots,n\}$, 
define $(L_n)_{A,B} = \one_{\{|A|=|B|+1 \wedge B \subset A\}}$ and $R_n = (L_n)^{T}$.
\ignore{
\[
	(L_n)_{S,T} = 
	\left\{\begin{array}{ll}
	1 & \mbox{ if $|S|=|T|+1$ and $T \subset S$ } \\
	0 & \mbox{ otherwise }
	\end{array}\right.
\]
and
\[
	(R_n)_{S,T} = 
	\left\{\begin{array}{ll}
	1 & \mbox{ if $|T|=|S|+1$ and $S \subset T$ } \\
	0 & \mbox{ otherwise }
	\end{array}\right.
\]
}
Now, let $H_n = R_n L_n - L_n R_n$. 
Notice that $A(Q_n)=L_n+R_n$.
\ignore{
Note 
$L_{n+1} = I_2 \otimes L_n + L_1 \otimes I_{2^n}$ and $R_{n+1} = I_2 \otimes R_n + R_1 \otimes I_{2^n}$,
where $L_1 = \begin{pmatrix} 0 & 0 \\ 1 & 0 \end{pmatrix}$ and
$R_1 = \begin{pmatrix} 0 & 1 \\ 0 & 0 \end{pmatrix}$.
}
Inductively, it can be shown that
\[
	[R_n,L_n] = H_n,
	\ \hspace{.1in} \
	[H_n,R_n] = 2R_n,
	\ \hspace{.1in} \
	[H_n,L_n] = -2L_n.
\]
\ignore{
Recall that the standard basis of $\mysl{2}{\CC}$ are given by
\[
	H = \begin{pmatrix} 1 & 0 \\ 0 & -1 \end{pmatrix},
	\ \hspace{.1in} \
	X = \begin{pmatrix} 0 & 1 \\ 0 & 0 \end{pmatrix}.
	\ \hspace{.1in} \
	Y = \begin{pmatrix} 0 & 0 \\ 1 & 0 \end{pmatrix},
\]
}
Thus, the homomorphism $\rho:\mysl{2}{\CC} \rightarrow \myend(V_n)$, with $|V_n| = 2^n$, 
where $\rho(H) = H_n$, $\rho(X) = R_n$, and $\rho(Y) = L_n$, is a representation of the 
Lie algebra $\mysl{2}{\CC}$ (where $H,X,Y \in \Mat_2(\CC)$ are the standard basis of $\mysl{2}{\CC}$).
This allows us to employ the following standard Lie theoretic argument (see \cite{hall,fh}).

The eigenvalues of $H_n$ are $\lambda=-n,-n+2,\ldots,n-2,n$. Let $W_\lambda$ be the eigenspace of $H_n$
corresponding to eigenvalue $\lambda$. By the properties of $\rho$ as a representation of $\mysl{2}{\CC}$, 
if $z \in W_\lambda$, then $R_n z \in W_{\lambda+2}$ (or $0$ if $\lambda = n$) 
and $L_n z \in W_{\lambda-2}$ (or $0$ if $\lambda = -n$).
Now, let $z$ be an eigenvector in the maximum eigenspace $W_n$.
Consider the invariant subspace $K_0$ spanned by the following set of orthogonal eigenvectors of $H_n$.
Let $z_k = L_n^k z$, $k=0,1,\ldots,n$, and set
\[
	K_0 = \spn\{z, L_n z, L_n^2 z, \ldots, L_n^n z\}.
\]
By definition, the action of $L_n$ on $K_0$ is given by $\ip{z_j}{L_n z_k} = \one_{\{j=k+1\}}$,
while the action of $R_n$ on $K_0$ is given by $\ip{z_j}{R_n z_k} = k(n-k)\one_{\{j=k-1\}}$.
Therefore, the action of $L_n+R_n$ on $K_0$ is similar to $A(Q_n/\pi)$ (a Krawtchouk chain of length $n+1$). 
Next, we repeat the argument on $K_0^\perp$ to extract additional Krawtchouk 
chains. This shows that the entire space $V_n$ can be decomposed into the invariant subspaces 
(or walk modules) as follows:
\[
	V_n = \bigoplus_{\ell \ge 0} K_\ell.
\]
As the action of $L_n+R_n$ (or the adjacency operator of $Q_n$) on each invariant subspace $K_\ell$ 
is similar to $A(Q_d/\pi)$, for some $d \le n$, perfect state transfer occur in all of these Krawtchouk chains.
\end{proof}
\end{theorem}

\subsection{Dual-Rail Encoding}

\newcommand{\WP}{\widetilde{P}}
\newcommand{\bket}[1]{e(#1)}
\newcommand{\bbra}[1]{e(#1)^T}

We describe a simple idea for constructing perfect state transfer in graphs with tails
based on the dual-rail encoding in spin chains (see Burgarth and Bose \cite{bb05},
and Kay \cite{k10} for a relevant survey).
Recall that the notation $\bket{x}$ denotes the unit vector corresponding to vertex $x$.

\begin{theorem} \label{thm:double-cover}
Let $G$ be a graph with perfect state transfer between vertices $u$ and $v$ at time $\tau$.
Let $H$ be the rooted product $P_3^{\mathcal{Y}}$ where $\mathcal{Y} = \{G_0,P_\infty,G_1\}$
with $G_0$ and $G_1$ represent two identical copies of $G$.
Then, $H$ has perfect state transfer between the states 
$\frac{1}{\sqrt{2}}(\bket{u_0}-\bket{u_1})$ and
$\frac{1}{\sqrt{2}}(\bket{v_0}-\bket{v_1})$
at time $\tau$, where $u_0,v_0$ and $u_1,v_1$ are vertices which correspond to the copies 
of $u,v$ in $G_0,G_1$, respectively.

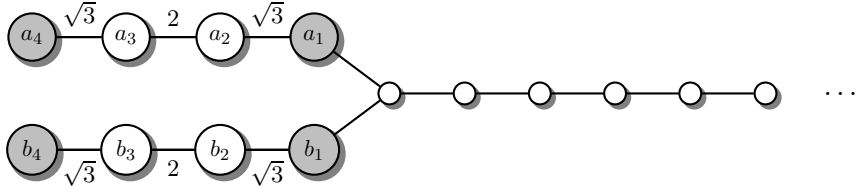
\begin{figure}[t]
\begin{center}
\begin{tikzpicture}[
    main node/.style={circle,draw,font=\bfseries}, main edge/.style={-,>=stealth'},
    scale=0.5,
    stone/.style={},
    black-stone/.style={black!80},
    black-highlight/.style={outer color=black!80, inner color=black!30},
    black-number/.style={white},
    white-stone/.style={white!70!black},
    white-highlight/.style={outer color=white!70!black, inner color=white},
    white-number/.style={black}]
\tikzset{every loop/.style={thick, min distance=17mm, in=45, out=135}}


\tikzstyle{every node}=[draw, thick, shape=circle, circular drop shadow, fill={gray!50}];
\path (-4.5,+1.5) node (a4) [scale=0.8] {$a_4$};
\path (-4.5,-1.5) node (b4) [scale=0.8] {$b_4$};

\tikzstyle{every node}=[draw, thick, shape=circle, circular drop shadow, fill={white}];
\path (-2.0,+1.5) node (a3) [scale=0.8] {$a_3$};
\path (-2.0,-1.5) node (b3) [scale=0.8] {$b_3$};

\tikzstyle{every node}=[draw, thick, shape=circle, circular drop shadow, fill={white}];
\path (+0.5,+1.5) node (a2) [scale=0.8] {$a_2$};
\path (+0.5,-1.5) node (b2) [scale=0.8] {$b_2$};

\tikzstyle{every node}=[draw, thick, shape=circle, circular drop shadow, fill={gray!50}];
\path (3.0,+1.5) node (a1) [scale=0.8] {$a_1$};
\path (3.0,-1.5) node (b1) [scale=0.8] {$b_1$};

\tikzstyle{every node}=[draw, thick, shape=circle, circular drop shadow, fill={white}];
\path (+5,0) node (q1) [scale=0.8] {};

\path (+7.0,0) node (pa) [scale=0.8] {};
\path (+9.0,0) node (pb) [scale=0.8] {};
\path (+11.0,0) node (pc) [scale=0.8] {};
\path (+13.0,0) node (pd) [scale=0.8] {};
\path (+15.0,0) node (pe) [scale=0.8] {};

\draw[thick]
    (a1) -- (a2) -- (a3) -- (a4);
\draw[thick]
    (b1) -- (b2) -- (b3) -- (b4);

\draw[thick]
    (q1) -- (a1)
	(q1) -- (b1)
	(q1) -- (pa);

\draw[thick]
    (pa) -- (pb) -- (pc) -- (pd) -- (pe);

\tikzstyle{every node}=[];
\node at (+17,0) {$\ldots$};

\path[-,draw,thick]
	(a1) edge node[above] {\mbox{\footnotesize $\sqrt{3}$}} (a2)
	(a2) edge node[above] {\mbox{\footnotesize $2$}} (a3)
	(a3) edge node[above] {\mbox{\footnotesize $\sqrt{3}$}} (a4);
\path[-,draw,thick]
	(b1) edge node[below] {\mbox{\footnotesize $\sqrt{3}$}} (b2)
	(b2) edge node[below] {\mbox{\footnotesize $2$}} (b3)
	(b3) edge node[below] {\mbox{\footnotesize $\sqrt{3}$}} (b4);
\end{tikzpicture}
\vspace{.1in}
\caption{A rooted product $P_3^{\mathcal{Y}}$ where $\mathcal{Y} = \{\widetilde{Q}_n,P_\infty,\widetilde{Q}_n\}$
and $\widetilde{Q}_n$ is a Krawtchouk chain (quotient of the $n$-cube).
{\em Perfect} pair state tranfer occurs between $\frac{1}{\sqrt{2}}(\bket{a_1} - \bket{b_1})$ 
and $\frac{1}{\sqrt{2}}(\bket{a_n}-\bket{b_n})$ at time $\pi/2$.
}
\label{fig:dual-tail}
\end{center}
\end{figure}

\begin{proof}
Let $(b,w)$ denote the vertices in the two copies of $G$, where $b \in \{0,1\}$ and $w \in V(G)$.
Let $\eket{0} = \begin{pmatrix} 1 \\ 0 \end{pmatrix}$ 
and $\eket{1} = \begin{pmatrix} 0 \\ 1 \end{pmatrix}$ be the standard two-dimensional unit vectors.
We show there is perfect state transfer in $H$ between the states
$\frac{1}{\sqrt{2}}(\eket{0}-\eket{1}) \otimes \bket{u}$ 
and
$\frac{1}{\sqrt{2}}(\eket{0}-\eket{1}) \otimes \bket{v}$
at time $\tau$.

Let $q_0$ be the middle vertex of $P_3$ in the rooted product $H = P_3^{\{G_0,P_\infty,G_1\}}$.
Also, let $r \in V(G)$ be the designated root vertex of $G$. 
Note that $A(H)$ is given by
\[
	A(H) =
	\begin{pmatrix}
	A(G)    & O       & \bket{r} &         \\
	O       & A(G)    & \bket{r} &         \\
	\bbra{r} & \bbra{r} & 0       & \ebra{1} \\
	        &         & \eket{1} & J_0
	\end{pmatrix}
\]
where the third row and column correspond to vertex $q_0$.
Consider the subspaces $W^{-}$ and $K$ defined as
\[
	W^{-} = \spn\left\{\frac{1}{\sqrt{2}}(\eket{0} - \eket{1}) \otimes \bket{w} : w \in V(G)\right\}
\]
and
\[
	K = \spn\left\{A(H)^m\bket{q_0} : m=0,1,\ldots\right\}.
\]

\begin{claim}
$W^{-}$ is orthogonal to $K$.

\begin{proof}
Let $W^+$ be a subspace defined as
\[
	W^+ = \spn\left\{\frac{1}{\sqrt{2}}(\eket{0} + \eket{1}) \otimes \bket{w} : w \in V(G)\right\}.
\]
Suppose the vertex set of $P_\infty$ is denoted as $\{q_n : n \ge 0\}$. 
Let $W^\infty$ be defined as $W^\infty = \spn\{\bket{q_n} : n \ge 0\}$.
Notice that 
\[
	A(H)\bket{q_0} = \bket{q_1} + \frac{1}{\sqrt{2}}(\eket{0}+\eket{1}) \otimes \bket{r}.
\]
Therefore, we see that
\[
	K = W^+ + W^\infty.
\]
Each vector in $W^{-}$ is orthogonal to any vector in $W^{+}$ since 
$\frac{1}{\sqrt{2}}(\eket{0} \pm \eket{1})$
form an orthogonal pair. Moreover, each vector in $W^{-}$ is orthogonal to any $\bket{q_n}$ since the
vertices of $G_0,G_1$ and $P_\infty$ are disjoint from each other.
\end{proof}
\end{claim}

\begin{claim}
$W^{-}$ is an invariant subspace of $A(H)$ and
the action of $A(H)$ restricted to $W^{-}$ is equivalent to $A(G)$.

\begin{proof}
The action of $A(H)$ on $\frac{1}{\sqrt{2}}(\eket{0}-\eket{1}) \otimes \bket{w}$ is given by
$I_2 \otimes A(G)$ (by the structure of the rooted product $H$). This shows $W^{-}$ is
an invariant subspace of $A(H)$ and that the action of $A(H)$ is equivalent to $A(G)$.
\end{proof}
\end{claim}

The above two claims show that for some unitary matrix $U$ we have
\[
	U^{-1} A(H) U = 
	\begin{pmatrix}
	A(G_n) & O \\
	O & \restr{A(H)}{K}
	\end{pmatrix}.
\]
This shows that there is perfect state transfer between
$\frac{1}{\sqrt{2}}(\eket{0}-\eket{1}) \otimes \ket{u}$
and
$\frac{1}{\sqrt{2}}(\eket{0}-\eket{1}) \otimes \ket{v}$
at time $\tau$.
\end{proof}
\end{theorem}

We remark that the construction used in Theorem \ref{thm:double-cover} applies also to 
the graph $G \Box K_2$ whose adjacency matrix is $I_2 \otimes A(G) + A(K_2) \otimes I$
(that is, trading the zero matrix for the identity matrix).

\section{Conclusions}

This work was motivated by the following question: can state transfer occur in a finite graph which is
connected to an infinite graph? We show that the answer is affirmative in graphs with tails. 
As such, our work was influenced by Golinskii \cite{g16} who studied the spectra of graphs with tails.
More surprisingly, we found that {\em perfect} state transfer is possible in graphs with tails.
Our main observation is that state transfer can occur in the finite invariant subspace obtained
from the decoupling theorem for eventually-free Jacobi matrices. 
It would be interesting to investigate state transfer (or lack thereof) in the infinite invariant
subspace. In fact, it is possible to replace the infinite tail with an arbitrary (but suitably defined)
finite graph. We leave this for future work.

We conclude with the following open questions.
\begin{enumerate}
\item The original motivation for this work is to investigate if spatial search works on infinite graphs.
	Although this question remains open, the corresponding problem for discrete-time quantum walk 
	was studied by Konno \etal \cite{kss21}.

\item Are there deeper connections lurking between the dark subspaces studied here and the non-primary modules 
	of the Terwilliger algebra? 

\item Corollary \ref{cor:random} shows $\widehat{G}(n,1/2)$ has no dark subspace. 
	But, what about $G(n,1/2)$?

\item What information about the dark (protected) subspace is revealed by scattering 
	(as employed by Farhi \etal \cite{fgg08})?

\end{enumerate}

\section*{Acknowledgments}

Work partly done while the authors were participants of the workshop
``Graph Theory, Algebraic Combinatorics, and Mathematical Physics''
at Centre de Recherches Math\'{e}matiques (CRM), Universit\'{e} de Montr\'{e}al.
C.T.\ would like to thank CRM for its hospitality and support during his sabbatical visit.
W.X.\ was supported by NSF travel grant DMS-2212755.
We thank Chris Godsil for helpful discussions.

\bibliographystyle{plain}

\begin{thebibliography}{99}

\bibitem{ag}
N.I. Akhiezer, I.M. Glazman.
\newblock {\em Theory of Linear Operators in Hilbert Space}.
\newblock Dover, 1993.

\bibitem{b03}
S. Bose.
\newblock {Quantum communication through an unmodulated spin chain}.
\newblock {\em Physical Review Letters} {\bf 91}(20):207901, 2003.

\bibitem{bb05}
D. Burgarth, S. Bose.
\newblock {Conclusive and arbitrarily perfect quantum state transfer using parallel spin chain channels}.
\newblock {\em Physical Review A} {\bf 71}:052315, 2005.

\bibitem{c09}
A. Childs.
\newblock {Universal Computation by Quantum Walk}.
\newblock {\em Physical Review Letters} {\bf 102}:180501, 2009.

\bibitem{cddekl05}
M. Christandl, N. Datta, T. Dorlas, A. Ekert, A. Kay, A. Landahl.
\newblock {Perfect state transfer of arbitrary states in quantum spin networks}.
\newblock {\em Physical Review A} {\bf 71}:032312, 2005.

\bibitem{cfghst14}
S. Cameron, S. Fehrenbach, L. Granger, O. Hennigh, S. Shreshtha, C. Tamon.
\newblock {Universal State Transfer on Graphs}.
\newblock {\em Linear Algebra and Its Applications} {\bf 455}:115-142, 2014.

\bibitem{cg20}
Q. Chen, C. Godsil.
\newblock {Pair state transfer}.
\newblock {\em Quantum Information Processing} {\bf 19}, 321, 2020.

\bibitem{cmf16}
X. Chen, R. Mereau, D. Feder.
\newblock {Asymptotically perfect efficient quantum state transfer across uniform chains with two impurities}.
\newblock {\em Physical Review A} {\bf 93}:012343, 2016.

\bibitem{fgg08}
E. Farhi, J. Goldstone, S. Gutmann.
\newblock {A Quantum Algorithm for the Hamiltonian NAND Tree}.
\newblock {\em Theory of Computing} {\bf 4}, Article 8, p169-190, 2008.

\bibitem{fh}
W. Fulton, J. Harris.
\newblock {\em Representation Theory: A First Course}.
\newblock Springer, 1991.

\bibitem{godsil-assoc}
C. Godsil.
\newblock {\em Association Schemes}.
\newblock unpublished notes, 2010.

\bibitem{g12}
C. Godsil.
\newblock {Controllable Subsets in Graphs}.
\newblock {\em Annals of Combinatorics} {\bf 16}:733-744, 2012.

\bibitem{g17}
C. Godsil.
\newblock {Sedentary quantum walks}.
\newblock arxiv:1710.11192 [math.CO].

\bibitem{gm78}
C. Godsil, B. McKay.
\newblock {A new graph product and its spectrum}.
\newblock {\em Bulletin of the Australian Mathematical Society} {\bf 18}:21-28, 1978.

\bibitem{gm88}
C. Godsil, B. Mohar.
\newblock {Walk Generating Functions and Spectral Measures of Infinite Graphs}.
\newblock {\em Linear Algebra and Its Applications} {\bf 107}:191-206, 1988.

\bibitem{gr}
C. Godsil, G. Royle.
\newblock {\em Algebraic Graph Theory}.
\newblock {Springer}, 2001.

\bibitem{g16}
L. Golinskii.
\newblock {Spectra of infinite graphs with tails}.
\newblock {\em Linear and Multilinear Algebra} {\bf 64}(11):2270-2296, 2016.

\bibitem{hall}
B. Hall.
\newblock {\em Lie Groups, Lie Algebras, and Representations}, 2nd edition.
\newblock Springer, 2015.

\bibitem{hj13}
R. Horn, C. Johnson.
\newblock {\em Matrix Analysis}, 2nd edition.
\newblock Cambridge, 2013.

\bibitem{jlr}
S.~Janson, T.~\L{}uczak, and A.~Ruci\'{n}ski.
\newblock {\em Random Graphs}.
\newblock Wiley, 2000.

\bibitem{k10}
A. Kay.
\newblock {Perfect, Efficient, State Transfer and Its Applications as a Constructive Tool}.
\newblock {\em International Journal of Quantum Information} {\bf 8}(4):641-676, 2010.

\bibitem{kss21}
N. Konno, E. Segawa, M. \v{S}tefa\v{n}\'{a}k.
\newblock Relation between quantum walks with tails and quantum walks with sinks on finite graphs.
\newblock arXiv:2105.03111 [math-ph].

\bibitem{ldsc18}
T. Le, L. Donati, S. Severini, F. Caruso.
\newblock {How to suppress dark states in quantum networks and bio-engineered structures}.
\newblock {\em Journal of Physics A: Mathematical and Theoretical} {\bf 51}(36):365306, 2018.

\bibitem{m82}
B. Mohar.
\newblock {The Spectrum of an Infinite Graph}.
\newblock {\em Linear Algebra and Its Applications} {\bf 48}:245-256, 1982.

\bibitem{mw89}
B. Mohar, W. Woess.
\newblock {A Survey on Spectra of Infinite Graphs}.
\newblock {\em Bulletin of London Mathematical Society} {\bf 21}:209-234, 1989.

\bibitem{ot16}
S. O'Rourke, B. Touri.
\newblock {On a Conjecture of Godsil Concerning Controllable Random Graphs}.
\newblock {\em SIAM Journal on Control and Optimization} {\bf 54}(6):3347-3378, 2016.

\bibitem{rudin}
W. Rudin.
\newblock {\em Functional Analysis}, 2nd edition.
\newblock McGraw-Hill, 1991.

\bibitem{terwilliger}
P. Terwilliger.
\newblock {Distance-regular graphs, the subconstituent algebra, and the Q-polynomial property}.
\newblock arXiv:2207.07747 [math.CO].

\end{thebibliography}

\end{document}